\documentclass[prd,floatfix,onecolumn,amsmath,amssymb]{revtex4-2}
\usepackage{graphicx,color,dcolumn,booktabs,bm}
\usepackage{subfigure}
\usepackage{amssymb}
\usepackage{longtable}
\usepackage{indentfirst}
\usepackage{epsfig,gensymb,siunitx}
\usepackage{feynmf}   
\usepackage{epstopdf}   
\usepackage{slashed}  
\usepackage{cases}
\usepackage{xcolor}
\definecolor{maroon}{RGB}{139,25,150}
\usepackage{multirow}
\usepackage{float}
\usepackage{pgf}

\usepackage{graphicx,color,dcolumn,booktabs,bm}
\usepackage[colorlinks, citecolor=blue,anchorcolor=red,menucolor=red, linkcolor=red,filecolor=red,runcolor=red,urlcolor=blue,frenchlinks=red, urlcolor=blue]{hyperref}
\usepackage{orcidlink}

\begin{document}
	
	\preprint{}

\title{Prompt diphoton production compared to measurements at 13 TeV in $k_t$-factorization: A comparative analysis of unintegrated PDF models}

\author{ CHROMA \footnote{Computational Hadronic Research for Observable Modeling and Analysis} Collaboration:\\
	R.~Kord Valeshabadi$^{1}$\orcidlink{0000-0001-5326-2459}}
	

\author{S.~Rezaie$^{1}$\orcidlink{0000-0002-1852-1619}}

\author{K.~Azizi$^{2,3,1}$\orcidlink{0000-0003-3741-2167}}
\email{kazem.azizi@ut.ac.ir}

\affiliation{
	$^{1}$School of Particles and Accelerators, Institute for Research in Fundamental Sciences (IPM), P.O. Box 19568-36681, Tehran, Iran\\
	$^{2}$Department of Physics, University of Tehran, North Karegar Avenue, Tehran 14395-547, Iran\\
	$^{3}$Department of Physics, Dogus University, Dudullu-Umraniye, 34775
	Istanbul, Turkiye
}

\begin{abstract}
	We perform an in-depth comparative analysis of unintegrated parton distribution function (UPDF) models for isolated prompt diphoton production in proton–proton collisions at $\sqrt{s}=13$~TeV within the $k_t$-factorization framework. Predictions are obtained with three UPDF approaches: Parton Branching (PB), NLO-MRW, and Modified KMRW (MKMRW). Tree-level $q + \bar q\!\to\!\gamma +\gamma$, $q + \bar q\!\to\!\gamma + \gamma + g$, and $q + g\!\to\!\gamma +\gamma + q$ subprocesses are generated with \textsc{KaTie} using off-shell initial states; the loop-induced $g + g\!\to\!\gamma + \gamma$ channel is evaluated independently. We compare differential cross sections with ATLAS measurements across a broad set of observables, including the photon transverse momenta, diphoton invariant mass and transverse momentum, the Collins–Soper angle, acoplanarity, $\phi^*_\eta$, and a transverse thrust–related variable. This comparative study quantifies the impact of the UPDF choice on the diphoton spectra. We find that the PB model provides the most consistent agreement over all distributions, whereas NLO-MRW tends to overshoot in regions correlated with larger factorization scales and MKMRW generally undershoots due to stronger Sudakov suppression. With standard scale variations, our results indicate that $k_t$-factorization with PB UPDFs can accurately describe diphoton production, while fixed-order collinear predictions typically require higher-order corrections together with parton-shower effects to achieve a comparable description.
\end{abstract}


\maketitle

\section{Introduction}

The production of isolated photon pairs (diphotons) in high-energy proton–proton collisions is of significant phenomenological and theoretical interest~\cite{2gnnlo,Gavardi:2022ixt,Campbell:2016lzl,Lipatov:2012td,Nefedov:2015ara,Modarres_2photon}. This process constitutes a principal and irreducible background in searches for new heavy neutral resonances decaying into two photons, such as the Standard Model Higgs boson~\cite{ATLAS:2012yve,CMS:2012qbp} and its possible extensions~\cite{ATLAS:2014jdv}. Moreover, since photons are largely insensitive to hadronization effects and can be measured with high precision, diphoton production provides a particularly clean environment for testing perturbative Quantum Chromodynamics (pQCD), both in the collinear~\cite{diphox,resbos,Campbell:2016lzl,Gavardi:2022ixt,Catani:2018krb} and $k_t$-factorization frameworks~\cite{Lipatov:2012td,Modarres_2photon,Nefedov:2015ara}.

Diphoton production has been extensively studied in various experiments across different kinematic regions and center-of-mass energies~\cite{CMS:2011xtn,CMS:2014mvm,CDF:2012ool,ATLAS:2011gau,ATLAS:2012fgo,ATLAS:2021mbt}. The challenge of achieving satisfactory agreement between theoretical predictions and experimental data, particularly through the inclusion of higher-order subprocesses, has been investigated in numerous studies within the collinear factorization framework~\cite{2gnnlo,NNLOJet,Campbell:2016lzl,Gavardi:2022ixt}. This challenge becomes even more pronounced at higher energies, where next-to-leading order (NLO) calculations such as \textsc{Diphox}~\cite{diphox} and \textsc{Resbos}~\cite{resbos} fail to fully reproduce the observed data. 

More recently, ATLAS measurements of diphoton production at $\sqrt{s}=13$~TeV~\cite{ATLAS:2021mbt} have shown that even next-to-next-to-leading order (NNLO) predictions from \textsc{NNLOJet}~\cite{NNLOJet} are insufficient to completely describe the measured cross sections. The study concluded that, to accurately model diphoton production within the collinear factorization framework, one must include not only higher-order perturbative corrections (to describe regions of high $p_T^{\gamma\gamma}$) but also parton-shower effects (essential in the low-$p_T^{\gamma\gamma}$ domain). These observations highlight the necessity of theoretical frameworks capable of simultaneously incorporating both resummation and higher-order QCD effects.

The $k_t$-factorization approach has shown great promise in addressing these challenges~\cite{Modarres_2photon,Lipatov:2012td,Nefedov:2015ara}, offering a more flexible description of initial-state parton dynamics by explicitly accounting for transverse momentum degrees of freedom. Unlike fixed-order collinear calculations, the $k_t$-factorization formalism inherently includes a portion of higher-order real emission effects through the transverse momentum of the incoming partons~\cite{smallXReview1}. This makes it particularly effective for describing observables sensitive to soft and collinear radiation, such as the azimuthal decorrelation and transverse momentum of the diphoton system.

In general, the computation of cross sections in hadronic collisions is complicated by the non-perturbative nature of the hadronic structure. Factorization theorems provide a rigorous framework to separate short-distance (perturbative) and long-distance (non-perturbative) dynamics. In the collinear factorization approach, partons are assumed to carry only a longitudinal momentum fraction $x$ of the proton momentum, with their distributions described by collinear parton distribution functions (PDFs), $f_a(x,\mu^2)$, which evolve with the factorization scale $\mu$ according to the (DGLAP) Dokshitzer–Gribov–Lipatov–Altarelli–Parisi  evolution equations~\cite{DGLAP1,DGLAP2,DGLAP3}. 

However, at small $x$, the transverse momentum $k_t$ of partons becomes non-negligible, and hence the collinear approximation becomes inadequate. In this regime, unintegrated parton distribution functions (UPDFs), $f_a(x,k_t^2,\mu^2)$, are required. These distributions incorporate both the longitudinal momentum fraction and the intrinsic transverse momentum of partons. The $k_t$-factorization approach~\cite{Smallx1,Smallx2,Smallx3} naturally includes both intrinsic and dynamically generated transverse momenta of partons, offering an improved description of QCD dynamics in regions where multiple scales are relevant.

The evolution of UPDFs has been formulated in several theoretical frameworks. At small $x$, the (BFKL) Balitsky–Fadin–Kuraev–Lipatov~\cite{BFKL1,BFKL2} evolution equation is applicable, while the  Catani–Ciafaloni–Fiorani–Marchesini (CCFM)~\cite{CCFM1,CCFM2,CCFM3,CCFM4} evolution equation interpolates between small and large $x$ regimes. However, both evolution equations are limited to the gluon sector. To achieve a complete description of the partonic structure, DGLAP-based approaches such as the Kimber–Martin–Ryskin (KMR)~\cite{KMR}, Martin–Ryskin–Watt (MRW)~\cite{MRW}, and Parton Branching (PB)~\cite{PB1,PB2} methods have been developed. These models have been successfully applied to a variety of processes in hadronic collisions~\cite{Valeshabadi:2021twn,PBDrell,Lipatov:2016wgr,Chernyshev:2023qea,Chernyshev:2024qvq}. 

Unlike the PB approach, which uses a Monte Carlo solution of the DGLAP equations and keeps the transverse momentum of partons to obtain realistic UPDFs, the LO-MRW and KMR formalisms often fail to describe data in regions sensitive to large transverse momenta.
These approaches typically produce broader distributions that overshoot the experimental data in such regions~\cite{Valeshabadi:2021twn,Valeshabadi:2021spa,Guiot:2022psv}. It has been shown, however, that the NLO-MRW formalism mitigates this issue by introducing a hard virtual-ordering cutoff, which suppresses large contributions from high-$k_t$ regions~\cite{Valeshabadi:2021twn}. Additionally, MRW-based approaches generally do not satisfy the normalization condition, and specifically at the LO level their differential and integral formulations can yield inconsistent results~\cite{Guiot:2019vsm,Golec-Biernat:2018hqo,Valeshabadi:2021smo,Guiot:2022psv}. To overcome these limitations, a modified version of the MRW formalism, referred to as MKMRW, was recently proposed~\cite{Guiot:2022psv}. This approach improves upon MRW by redefining the Sudakov form factor and modifying the differential form of the LO-MRW equations.

We aim to test the predictive power of the $k_t$-factorization framework for diphoton production using the latest ATLAS data at $\sqrt{s}=13$~TeV~\cite{ATLAS:2021mbt}. This work is the first analysis of prompt diphoton production using the PB and MKMRW methods. We present differential cross section predictions employing several UPDF models, including NLO-MRW, MKMRW, and PB, to evaluate their consistency with experimental data and to gain insight into their underlying dynamical behavior. The relevant partonic subprocesses considered include $q + \bar{q} \to \gamma + \gamma$, $q +\bar{q} \to \gamma + \gamma + g$, $q + g \to \gamma + \gamma + q$, and $g + g \to \gamma + \gamma$. The first three subprocesses are generated using the \textsc{KaTie} parton-level event generator~\cite{KATIE}, which implements off-shell initial-state partons and is restricted to tree-level matrix elements. The loop-induced $g + g \to \gamma + \gamma$ contribution, which cannot be handled within \textsc{KaTie}, is computed independently. 

In our analysis, we investigate a range of kinematic observables sensitive to different dynamical regimes, including the transverse momenta of the individual photons ($p_{T,\gamma_1}$, $p_{T,\gamma_2}$), the invariant mass and transverse momentum of the diphoton system ($m_{\gamma\gamma}$ and $p_{T,\gamma\gamma}$), the Collins–Soper scattering angle $|\cos\theta^*_{\mathrm{(CS)}}|$, the angular variable $\phi^*_{\eta}$, the acoplanarity $\pi - \Delta\phi_{\gamma\gamma}$, and the transverse thrust-related variable $a_{T,\gamma\gamma}$. These observables, particularly those sensitive to soft radiation such as $\phi^*_{\eta}$ and $a_{T,\gamma\gamma}$, provide important tests of the $k_t$-factorization approach and its implementation through different UPDF models. 

The structure of this paper is as follows. Section~II outlines the $k_t$-factorization framework and  introduces the UPDFs used in this study. Section~III describes the numerical methods. Section~IV presents and discusses the results, and Section~V summarizes the main conclusions.

\section{$k_t$-factorization framework}
\label{sec:kt-fact}
As discussed in the Introduction, the $k_t$-factorization framework enables the calculation of cross sections by explicitly accounting for the transverse momentum of partons participating in the hard subprocess. In this approach, the hadronic cross section is expressed as a convolution of the UPDFs and the partonic cross section, formally written as:
\begin{equation}
	\label{eq:kt-cs}
	\sigma = \sum_{a,b=q,g} \int \frac{dx_{1}}{x_{1}} \frac{dx_{2}}{x_{2}} \, dk_{1t}^2 \, dk_{2t}^2 \,
	f_{a}(x_{1}, k_{1t}^2, \mu^2)\,
	f_{b}(x_{2}, k_{2t}^2, \mu^2)\,
	\hat{\sigma}^*_{ab}.
\end{equation}

In the above equation, $\hat{\sigma}^*_{ab}$ denotes the off-shell partonic cross section for incoming partons $a$ and $b$, which includes their transverse momenta. The UPDFs, denoted as $f_{a(b)}(x, k_t^2, \mu^2)$, are characterized by dependence on three parameters: the longitudinal momentum fraction $x$ of the parton with respect to its parent hadron, the transverse momentum $k_t$ of the parton, and the factorization scale $\mu$ at which they are computed. As discussed earlier, the UPDFs constitute one of the key ingredients of equation Eq.~\eqref{eq:kt-cs}, therefore, in the three next subsections, we introduce the specific UPDF formalisms employed in this work.

\subsection{The PB UPDFs method}

The PB methodology~\cite{PB1,PB2} establishes a structured procedure for the evolution of PDFs from an initial low-energy scale, at which they are parameterized, to the appropriate hard scale relevant to the physical process. This evolution proceeds according to the DGLAP equations. By assigning a physical interpretation to the evolution scale, the PB method enables the calculation of the parton transverse momentum during the evolution, thereby facilitating the construction of UPDFs~\cite{PB2}.

In this framework, the evolution is performed using next-to-leading order (NLO) DGLAP splitting functions with angular ordering applied. The evolution scale $\mu_i$ is related to the transverse momentum of the emitted parton according to:
\begin{equation}
	{\bf q}_{t,i}^2 = (1 - z_i)^2 \mu_i^2,
\end{equation}
where $z_i$ represents the longitudinal momentum fraction associated with the branching. The PB approach allows for the derivation of UPDFs for both quarks and gluons. Its numerical implementation employs an iterative Monte Carlo solution that samples the next branching scale and splitting variable, generating an exclusive cascade and providing both integrated PDFs and UPDFs. In the collinear limit, the PB solution agrees with conventional DGLAP evolution at the percent level over a wide $(x, \mu)$ range and reproduces HERA~I+II DIS data within the \textsc{xFitter} framework.

At the initial scale $\mu_0$, the UPDFs are assumed to factorize into a Gaussian dependence on the transverse momentum and a parameterized dependence on $x$ and $\mu_0$. In general, the PB UPDFs can be expressed as:
\begin{equation*}
	\label{eq:22}
	F_a(x, k_t^2, \mu^2) =
	\Delta_a(\mu^2)\, F_a(x, k_t^2, \mu_0^2)
	+ \sum_b \int
	\dfrac{d^2 \bm{q}'}{\pi |\bm{q}'|^2}
	\dfrac{\Delta_a(\mu^2)}{\Delta_a(|\bm{q}'|^2)}
	\Theta(\mu^2 - |\bm{q}'|^2)
	\Theta(|\bm{q}'|^2 - \mu_0^2)
\end{equation*}
\begin{equation}
	\int_x^{z_M} \dfrac{dz}{z}\,
	P_{ab}^{R}(\alpha_s, z)\,
	F_b\!\left(\dfrac{x}{z}, k_t^{\prime 2}, |\bm{q}'|^2\right),
\end{equation}
where $\bm{k}'_t = \bm{q}'(1 - z) + \bm{k}_t$. The splitting function $P_{ab}^{R}(\alpha_s, z)$ is decomposed into two components: one containing the soft-gluon emission singularity and another comprising logarithmic and analytic terms, as detailed in \cite{PB1,PB2}. The Sudakov form factor $\Delta_a(\mu^2)$ is defined as:
\begin{equation}
	\label{eq:23}
	\Delta_a(\mu^2) =
	\exp\!\bigg(
	-\sum_b \int_{\mu_0^2}^{\mu^2}
	\dfrac{d\mu^{\prime 2}}{\mu^{\prime 2}}
	\int_0^{z_M} dz\, z\, P_{ba}^{R}(\alpha_s, z)
	\bigg),
\end{equation}
where $z_M$ denotes the soft-gluon resolution scale separating real and virtual emissions.

Finally, the UPDFs satisfy the normalization condition
\begin{equation}
	\label{eq:normalization}
     \int_{0}^{\mu^2}d{k}_t^2\;
	F_a(x, {k}_t^2, \mu^2)
	= f_a(x, \mu^2),
\end{equation}
providing an internal consistency check and a direct connection to the standard collinear factorization framework.

\subsection{The MRW UPDFs method}
\label{LO-NLO-MRWSec}

The LO-MRW and NLO-MRW unintegrated parton distribution functions (UPDFs) are DGLAP-based transverse momentum dependent (TMD) formulations, similar to the PB approach, and enable the determination of UPDFs for both quarks and gluons. The general expression of the MRW UPDFs for (anti)quarks and gluons is given by~\cite{MRW}:
\begin{equation}
	\label{eq:MRW}
	F_a(x, k_t^2, \mu^2) = 
	T_a(k_t^2, \mu^2)\,
	\dfrac{\alpha_s(k_t^2)}{2 \pi k_t^2}\,
	\sum_{b=q,g}\int_x^1 dz\, P_{ab}(z)\,
	f_b\!\left(\dfrac{x}{z}, k_t^2\right)\,
	\Theta^{\delta^{ab}}(z_{\text{max}} - z),
\end{equation}
where $T_a(k_t^2, \mu^2)$ is the Sudakov form factor, $f_b(x/z, k_t^2)$ denotes the momentum-weighted parton density at leading order (LO), and $\Theta^{\delta^{ab}}(z_{\text{max}} - z)$ imposes a constraint on gluon emission when $a = b$.

The Sudakov form factor is defined as
\begin{equation*}
	T_a(k_t^2 \leq \mu^2, \mu^2) =
	\exp\!\left(
	-\int_{k_t^2}^{\mu^2}
	\dfrac{d\kappa_t^2}{\kappa_t^2}
	\dfrac{\alpha_s(\kappa_t^2)}{2\pi}
	\sum_{b=q,g}\int_0^1 d\xi\, \xi\, P_{ba}(\xi)
	\Theta^{\delta^{ab}}(\xi_{\text{max}} - \xi)
	\right),
\end{equation*}
\begin{equation}
	\label{eq:Sud}
	T_a(k_t^2 > \mu^2, \mu^2) = 1.
\end{equation}

The MRW UPDFs are constructed via the DGLAP evolution equations, thereby permitting the computation of UPDFs for both quarks and gluons. In this framework, partons undergo collinear evolution  within the proton under DGLAP dynamics up to the final step, at which a real emission generates the transverse momentum $k_t$ dependence. The partons then evolve to the factorization scale $\mu$ in the absence of additional real emissions, with the process governed by the Sudakov form factor.

Equation~\eqref{eq:MRW} is valid only for $k_t \geq \mu_0$, where $\mu_0 \simeq 1~\text{GeV}$, since the input PDFs $f_a(x, k_t^2)$ are undefined below this scale. To extend the MRW UPDFs to $k_t < \mu_0$, the normalization condition given in Eq.~\eqref{eq:normalization} can be applied, ensuring consistency with the collinear PDFs. For $k_t < \mu_0$, the following expression satisfies this condition:
\begin{equation}
	\label{eq:MRWLessMu0}
	\dfrac{1}{k_t^2}\, f_a(x, k_t^2 < \mu_0^2, \mu^2)
	= \dfrac{1}{\mu_0^2}\, f_a(x, \mu_0^2)\,
	T_a(\mu_0^2, \mu^2),
\end{equation}
which results in a constant distribution for $k_t < \mu_0$.

The LO-MRW UPDFs for (anti)quarks and gluons, along with their corresponding Sudakov form factors, are expressed as~\cite{MRW}:
\begin{equation}
	\label{eq:LOMRWQ}
	\begin{split}
		F_q(x, k_t^2, \mu^2) &=
		T_q(k_t^2, \mu^2)\,
		\dfrac{\alpha_{s}^{\text{LO}}(k_t^2)}{2\pi k_t^2}
		\int_x^1\!
		\Big[
		P_{qq}^{\text{LO}}(z)\,
		f_q^{\text{LO}}\!\left(\dfrac{x}{z}, k_t^2\right)
		\Theta(z_{\text{max}} - z)
		+ P_{qg}^{\text{LO}}(z)\,
		f_g^{\text{LO}}\!\left(\dfrac{x}{z}, k_t^2\right)
		\Big]\, dz,
	\end{split}
\end{equation}
\begin{equation}
	\label{eq:LOMRWG}
	\begin{split}
		F_g(x, k_t^2, \mu^2) &=
		T_g(k_t^2, \mu^2)\,
		\dfrac{\alpha_{s}^{\text{LO}}(k_t^2)}{2\pi k_t^2}
		\int_x^1\!
		\Big[
		P_{gg}^{\text{LO}}(z)\,
		f_g^{\text{LO}}\!\left(\dfrac{x}{z}, k_t^2\right)
		\Theta(z_{\text{max}} - z)
		+ \sum_q P_{gq}^{\text{LO}}(z)\,
		f_q^{\text{LO}}\!\left(\dfrac{x}{z}, k_t^2\right)
		\Big]\, dz,
	\end{split}
\end{equation}
with the corresponding Sudakov form factors:
\begin{equation}
	\label{eq:8}
	T_q(k_t^2, \mu^2) =
	\exp\!\left(
	-\int_{k_t^2}^{\mu^2}
	\dfrac{d\kappa_t^2}{\kappa_t^2}\,
	\dfrac{\alpha_{s}^{\text{LO}}(\kappa_t^2)}{2\pi}
	\int_0^1 P_{qq}^{\text{LO}}(\xi)\,
	\Theta(\xi_{\text{max}} - \xi)\, d\xi
	\right),
\end{equation}
\begin{equation}
	\label{eq:9}
	T_g(k_t^2, \mu^2) =
	\exp\!\left(
	-\int_{k_t^2}^{\mu^2}
	\dfrac{d\kappa_t^2}{\kappa_t^2}\,
	\dfrac{\alpha_{s}^{\text{LO}}(\kappa_t^2)}{2\pi}
	\int_0^1
	\Big[
	\xi\, P_{gg}^{\text{LO}}(\xi)\,
	\Theta(\xi_{\text{max}} - \xi)\,
	\Theta(\xi - \xi_{\text{min}})
	+ n_F\, P_{qg}^{\text{LO}}(\xi)
	\Big] d\xi
	\right),
\end{equation}
where $q = \{u, \bar{u}, d, \bar{d}, \ldots\}$ and $n_F$ is the number of active quark flavours.

The angular ordering condition (AOC) is applied in the final evolution step to determine the cutoffs on $z$ and $\xi$, which constrain gluon emission in Eqs.~\eqref{eq:MRW} and~\eqref{eq:Sud}. The cutoffs are given by:
\begin{equation}
	z_{\text{max}} = \frac{\mu}{\mu + k_t}, \qquad
	\xi_{\text{max}} = \frac{\mu}{\mu + \kappa_t},
\end{equation}
with $\xi_{\text{min}} = 1 - \xi_{\text{max}}$.

The LO-MRW formalism also admits a differential representation:
\begin{equation}
	\label{eq:15}
	F_a(x, k_t^2, \mu^2) =
	\dfrac{\partial}{\partial k_t^2}
	\big[ F_a(x, k_t^2)\, T_a(k_t^2, \mu^2) \big].
\end{equation}
However, as discussed in Refs.~\cite{Guiot:2019vsm,Golec-Biernat:2018hqo,Valeshabadi:2021smo,Guiot:2022psv}, the differential and integral formulations of these UPDFs are not equivalent. To obtain consistent results, one must employ cutoff-dependent PDFs rather than collinear ones and modify the differential form accordingly. It is also noted in Ref.~\cite{Guiot:2022psv} that the cutoff-dependent form does not necessarily improve agreement with experimental data. Furthermore, the LO-MRW formalism exhibits an unphysically large tail at high transverse momentum due to the freedom of partons to acquire $k_t > \mu$, leading to unreliable predictions in such regions. In this work, our calculations show that the LO-MRW approach does not yield satisfactory results; therefore, we do not present results corresponding to this UPDF model.

Despite these shortcomings, the MRW formalism has been extended to next-to-leading order (NLO-MRW), which provides improved agreement with experimental data, such as Drell–Yan~\cite{Valeshabadi:2021spa} and three-photon production~\cite{Valeshabadi:2021twn}. The NLO-MRW UPDFs are obtained by modifying the DGLAP scale and employing NLO splitting functions, as given in Eqs.~\eqref{eq:NLO-MRWQ} and~\eqref{eq:NLO-MRWG}. In this approach, the input PDF scale is taken as $k^2 = k_t^2 / (1 - z)$ rather than $k_t^2$. An additional strong ordering constraint (SOC), $\Theta(\mu^2 - k^2)$, is introduced to restrict the parton transverse momentum to $k_t^2 < \mu^2$. This constraint significantly affects the behaviour of the UPDFs in the regions of large $k_t$ ($k_t \simeq \mu$) and large $z$ ($z \simeq 1$), leading to a suppression of the UPDFs in these regimes. 

Martin \textit{et al.}~\cite{MRW} further demonstrate that the NLO corrections to the splitting functions are numerically small compared to the LO contributions, and that UPDFs obtained with NLO collinear PDFs but only LO splitting functions reproduce, to very good accuracy, those obtained with the full NLO splitting functions over a wide range of $x$, $k_t$ and scales. On this basis they advocate a simplified NLO prescription in which one keeps the NLO PDFs, kinematics and Sudakov form factors, but uses LO splitting functions in the final evolution step. We therefore adopt this simplified NLO-MRW prescription, and the NLO-MRW formalism can be expressed as:
\begin{equation*}
	F_q(x, k_t^2, \mu^2) =
	\int_x^1
	T_q(k^2, \mu^2)\,
	\dfrac{\alpha_{s}^{\text{NLO}}(k^2)}{2\pi k_t^2}
	\Big[
	P_{qq}^{\text{LO}}(z)\,
	f_q^{\text{NLO}}\!\left(\dfrac{x}{z}, k^2\right)
	\Theta(z_{\text{max}} - z)
\end{equation*}
\begin{equation}
	\label{eq:NLO-MRWQ}
	+ P_{qg}^{\text{LO}}(z)\,
	f_g^{\text{NLO}}\!\left(\dfrac{x}{z}, k^2\right)
	\Big]
	\Theta(\mu^2 - k^2)\, dz,
\end{equation}
\begin{equation*}
	F_g(x, k_t^2, \mu^2) =
	\int_x^1
	T_g(k^2, \mu^2)\,
	\dfrac{\alpha_{s}^{\text{NLO}}(k^2)}{2\pi k_t^2}
	\Big[
	P_{gg}^{\text{LO}}(z)\,
	f_g^{\text{NLO}}\!\left(\dfrac{x}{z}, k^2\right)
	\Theta(z_{\text{max}} - z)
\end{equation*}
\begin{equation}
	\label{eq:NLO-MRWG}
	+ \sum_q P_{gq}^{\text{LO}}(z)\,
	f_q^{\text{NLO}}\!\left(\dfrac{x}{z}, k^2\right)
	\Big]
	\Theta(\mu^2 - k^2)\, dz,
\end{equation}
with the corresponding Sudakov form factors:
\begin{equation}
	\label{eq:12}
	T_q(k^2, \mu^2) =
	\exp\!\left(
	-\int_{k^2}^{\mu^2}
	\dfrac{dp^2}{p^2}\,
	\dfrac{\alpha_{s}^{\text{NLO}}(p^2)}{2\pi}
	\int_0^1
	P_{qq}^{\text{LO}}(\xi)\,
	\Theta(\xi_{\text{max}} - \xi)\, d\xi
	\right),
\end{equation}
\begin{equation}
	\label{eq:13}
	T_g(k^2, \mu^2) =
	\exp\!\left(
	-\int_{k^2}^{\mu^2}
	\dfrac{dp^2}{p^2}\,
	\dfrac{\alpha_{s}^{\text{NLO}}(p^2)}{2\pi}
	\int_0^1
	\Big[
	P_{gg}^{\text{LO}}(\xi)\,
	\Theta(\xi_{\text{max}} - \xi)\,
	\Theta(\xi - \xi_{\text{min}})
	+ n_F\, P_{qg}^{\text{LO}}(\xi)
	\Big] d\xi
	\right).
\end{equation}

Since $k^2$ depends on $z$, both the strong coupling constant and the Sudakov form factors in Eqs.~\eqref{eq:NLO-MRWQ} and~\eqref{eq:NLO-MRWG} must be evaluated inside the $z$ integral. This dependence significantly increases the computational complexity of the NLO-MRW formalism. In this work, the angular ordering condition is adopted for $\xi_{\text{max}}$ and $z_{\text{max}}$ in the NLO-MRW UPDFs.

\subsection{The MKMRW UPDFs method}

In this section, we introduce the MKMRW approach to UPDFs, as proposed by Guiot~\cite{Guiot:2022psv}. This formalism addresses several limitations of the original angular-ordered (AO) KMRW UPDFs~\cite{KMR,MRW}, including the overestimation of heavy-meson cross sections, inconsistencies between the differential and integral formulations, and issues related to the normalization condition. The key innovation of the MKMRW method is a revised normalization scheme, in which the integration over the transverse momentum $k_t$ extends to infinity. This ensures that the entire high-$k_t$ tail of the distribution is constrained and remains consistent with the corresponding PDFs.

The standard KMRW UPDFs satisfy the normalization condition:
\begin{equation}
	\tilde{f}_a(x, \mu) = \int_0^{\mu^2} F_a(x, k_t, \mu)\, dk_t^2,
\end{equation}
where $\tilde{f}_a$ represents the momentum-weighted parton density. However, this definition leaves the region $k_t > \mu$ unconstrained, which can lead to phenomenological issues such as the overestimation of $D$-meson production cross sections in hadron–hadron collisions. The MKMRW approach adopts a PB–inspired normalization~\cite{PB1,PB2}, consistent with Eq.~\eqref{eq:normalization}, which integrates over all $k_t$ values. This modification removes the unconstrained tail, avoids divergences associated with the light-cone gauge~\cite{Guiot:2022psv}, and ensures full consistency between the UPDFs and the collinear PDFs across the entire transverse momentum range. At the same time, the approach preserves the angular-ordering coherence characteristic of the original KMRW framework. Comparisons with AO–KMRW and PB UPDFs show that MKMRW distributions are suppressed at high $k_t$, resulting in improved agreement with collinear PDFs when integrated up to infinity.

The MKMRW formalism begins with a differential definition:
\begin{equation}
	F_a(x, k_t, \mu) =
	\frac{\partial}{\partial k_t^2}
	\left[ T_a(k_t, \mu)\, \tilde{f}_a(x, \mu) \right],
\end{equation}
where the argument of the collinear PDF is fixed at $\mu$ to comply with DGLAP evolution and to avoid unphysical limits as $k_t \to 0$ or $k_t \to \infty$. The Sudakov form factor $T_a(k_t, \mu)$ is redefined as:
\begin{equation}
	\label{eq:MKMRWSud}
	T_a(k_t, \mu) =
	\exp\!\left(
	-\int_{k_t^2}^{\infty}
	\frac{dq^2}{q^2}\,
	\frac{\alpha_s(q^2)}{2\pi}
	\sum_b \int_0^{\Delta(q, \mu)} dz\, z\, \hat{P}_{ba}(z)
	\right),
\end{equation}
with the angular-ordering cutoff $\Delta(q, \mu) = \mu / (\mu + q)$. This definition guarantees that $T_a(\infty, \mu) = 1$ and $T_a(0, \mu) = 0$, thereby preventing unphysical values greater than unity and ensuring proper normalization.

By differentiating the above expression, one obtains the integral representation:
\begin{equation}
	F_a(x, k_t, \mu) =
	\frac{\alpha_s(k_t^2)}{2\pi k_t^2}\,
	T_a(k_t, \mu)\,
	\tilde{f}_a(x, \mu)\,
	\sum_b \int_0^{\Delta(k_t, \mu)} dz\, z\, \hat{P}_{ba}(z).
\end{equation}

\section{Numerical Methods and Experimental Cuts}
We compute the cross section for isolated prompt diphoton production at $\sqrt{s} = 13~\mathrm{TeV}$ within the $k_t$-factorization framework, accounting for both tree-level and one-loop contributions. The calculation is performed using the \textsc{KaTie} parton-level event generator~\cite{KATIE} with $n_f=5$, which we employ to evaluate the tree-level subprocesses:

\begin{enumerate}
	\item $q +\bar q\to\gamma+\gamma$,
	\item $q +\bar q\to\gamma+\gamma+g$,
	\item $q+g\to\gamma+\gamma+q$.
\end{enumerate}
The contribution from the $g + g \rightarrow \gamma + \gamma$ one-loop level subprocess is computed separately, as \textsc{KaTie} does not support loop-level amplitude calculations. Consequently, the $g + g \rightarrow \gamma + \gamma$ channel, which provides an essential component of the total cross section, is evaluated independently. Following the approach adopted in Refs.~\cite{Lipatov:2012td,Modarres_2photon}, we refrain from including the full off-shell one-loop amplitude. Instead, we restrict our treatment to the collinear amplitude derived in Ref.~\cite{Bern:2002jx}, while consistently incorporating the off-shell kinematics of the incoming gluons.

The computation of the partonic cross sections in \textsc{KaTie} requires an appropriately defined input configuration file. This file specifies the subprocess type, the input PDFs, the UPDFs, kinematic cuts, and other relevant parameters. In particular, the UPDFs constitute a central ingredient of the $k_t$-factorization framework. They can be supplied either as grid files following the \textsc{KaTie} standard format or through the TMDLib2 interface \cite{TMDLib}. In this analysis, we employ the PB-NLO-HERAI+II-2023-set2-qs=1.04 UPDF set for the PB UPDFs using TMDLib2. For the MKRMW and NLO-MRW UPDFs, we generate sets in a custom format compatible with \textsc{KaTie}, in which the UPDFs are obtained using the collinear PDF set \texttt{MSHT20nlo\_as120}~\cite{Bailey:2020ooq}. The generation of these UPDFs relies heavily on the PDFxTMDLib library~\cite{PDFxTMDLib}, which provides the required collinear PDF inputs. To ensure consistency, the collinear PDFs from PDFxTMDLib are cross-checked against the LHAPDF library~\cite{LHAPDF6}.

For the independent computation of the $g + g \rightarrow \gamma + \gamma$ subprocess, where \textsc{KaTie} is not utilized, the PDFxTMDLib library again serves as a core component of our numerical framework, providing the UPDF inputs consistently across all subprocesses. It should be noted that for the MKMRW UPDF model we have upper limit of the Sudakov form factor integration infinity, see Eq.~\eqref{eq:MKMRWSud}, and as discussed with the author of this model we set infinity equal to maximum of $\mu_F$  defined by a PDF set. The UPDFs employed in this study are publicly available at \url{https://pdfxtmdlib.org/downloads}. Also, for evaluating multi-dimensional integrals (see, for example, equation 28 of the reference \cite{Lipatov:2012td}), we use the VEGAS Monte Carlo algorithm \cite{Lepage:1977sw}.

To ensure reliable predictions, we adopt the following choices for scales and uncertainty estimation. For the calculation of the cross section, we choose the factorization and renormalization scales as $\mu_{F,R}^{\mathrm{central}} = \sqrt{p_{T,\gamma\gamma}^{2} + m_{\gamma\gamma}^2}$. The scale uncertainty is estimated by repeating the calculation with the varied scales $\mu_{F,R}^{\mathrm{upper}}=2\mu_{F,R}^{\mathrm{central}}$ and $\mu_{F,R}^{\mathrm{lower}}=0.5\mu_{F,R}^{\mathrm{central}}$. Additionally, to avoid double counting between subprocesses 1 and 2, we follow the procedure introduced in~\cite{Maciula:2019izq}. The goal of this approach is to limit the hard final state gluons to subprocess 2 and the soft ones (coming from the UPDF) to subprocess 1.To impose this constraint, for subprocess 1, the UPDFs in this subprocess are restricted such that $k_t \leq \mu$. Furthermore, for subprocess 2, the transverse momentum of the final state gluon satisfies $k_{T,g} > \max\{k_{T,A}, k_{T,B}\}$, where $k_{T,A}$ and $k_{T,B}$ are the transverse momenta of the incoming partons along the $\hat{z}$ and $-\hat{z}$ directions, respectively.

Finally, the experimental cuts of the ATLAS experiment \cite{ATLAS:2021mbt} are imposed as follows:
\begin{enumerate}
	\item Photons must be separated from each other by $\Delta R_{ij} > 0.4$, where $\Delta R_{ij}=\sqrt{(\eta_i - \eta_j)^2+(\phi_i-\phi_j)^2}$, and $i$ and $j$ are the first and second final state photons.
	\item Transverse momenta of the two photons are: $p_T^{\gamma_1} > 40\; \mathrm{GeV}$, $p_T^{\gamma_2} > 30\; \mathrm{GeV}$.
	\item Two photons must have pseudo-rapidities $|\eta^{\gamma}| \leq 2.37$ or $1.37 \leq |\eta^{\gamma}| \leq 1.52$.
	\item Instead of the standard isolation cone implemented in the experiment, we impose a smooth isolation cone \cite{Frixione:1998jh} with benefits that on one side it regularizes the photon collinear divergence and on the other side it suppresses the fragmentation contribution \cite{Valeshabadi:2021twn}.
	The smooth isolation cone enforces the transverse energies of particles around the distance $\Delta R \leq R_0$ from each photon to be:
	\begin{equation}
		E_T^{\mathrm{iso}}(\Delta R) < \chi(\Delta R),
	\end{equation}
	where $E_T^{\mathrm{iso}}$ is the sum of transverse energies of the particles around the photon at a distance less than $\Delta R$. The function $\chi(\Delta R)$ is defined as:
	\begin{equation}
		\chi(\Delta R) = E_{T,\gamma}\epsilon_{\gamma}\left(\frac{1-\cos\Delta R}{1-\cos R_0}\right)^n.
	\end{equation}
Following the setup of the Sherpa~2.2 and NNLOJET predictions compared to the same ATLAS measurement~\cite{ATLAS:2021mbt}, we choose
\begin{equation}
	R_0 = 0.1,\qquad n = 2,\qquad \epsilon_{\gamma} = 0.1.
\end{equation}
It should be noted that in the ATLAS fiducial phase space a standard cone of radius $\Delta R = 0.2$ is used with an isolation requirement $E_{T,\gamma}^{\mathrm{iso},0.2} < 0.09\,p_{T,\gamma}$~\cite{ATLAS:2021mbt}. For such tight isolation, dedicated studies have shown that smooth-cone and standard-cone prescriptions yield very similar predictions for LHC diphoton observables~\cite{Cieri:2015wwa}, within theoretical uncertainties, while smooth-cone isolation retains the advantage of suppressing the fragmentation component.

\end{enumerate}

These methods and cuts enable a direct comparison with ATLAS measurements, as detailed in the subsequent sections.
\section{Results and Discussion}

In this section, we present the results of the differential cross section calculations for diphoton production in proton–proton collisions at $\sqrt{s} = 13~\mathrm{TeV}$ using various UPDF models within the $k_{t}$-factorization framework. 
The kinematical cuts imposed on the calculations follow those applied in the ATLAS measurement~\cite{ATLAS:2021mbt}. 

We first introduce the kinematical variables investigated in this work, which are defined consistently with Ref.~\cite{ATLAS:2021mbt}:

\begin{figure}[htbp]
	\centering
	\includegraphics[width=16cm, height=7.1cm]{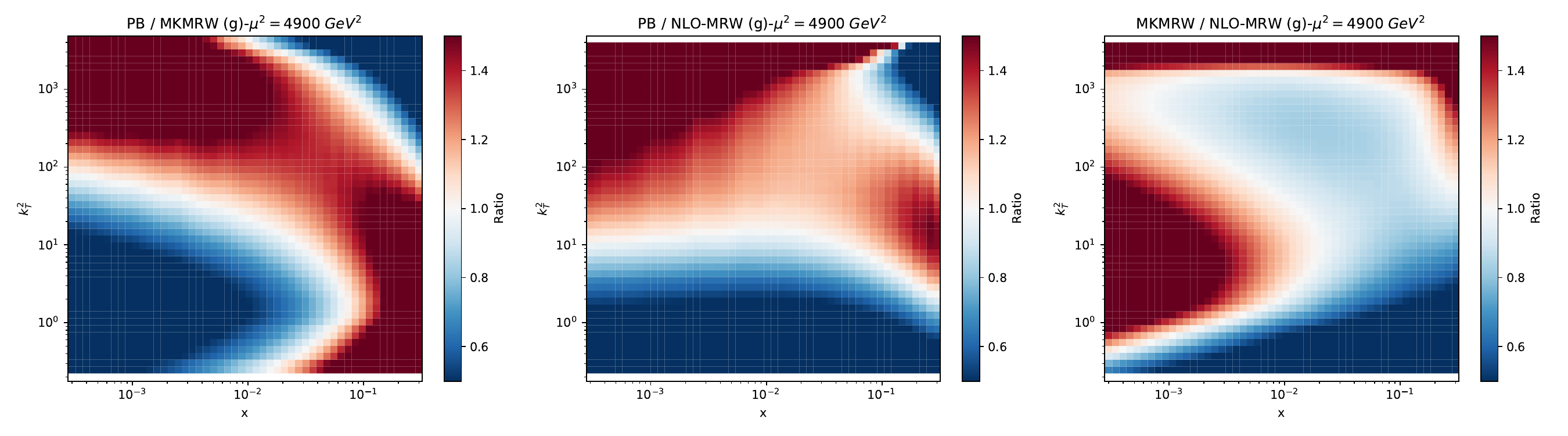}
	\includegraphics[width=16cm, height=7.1cm]{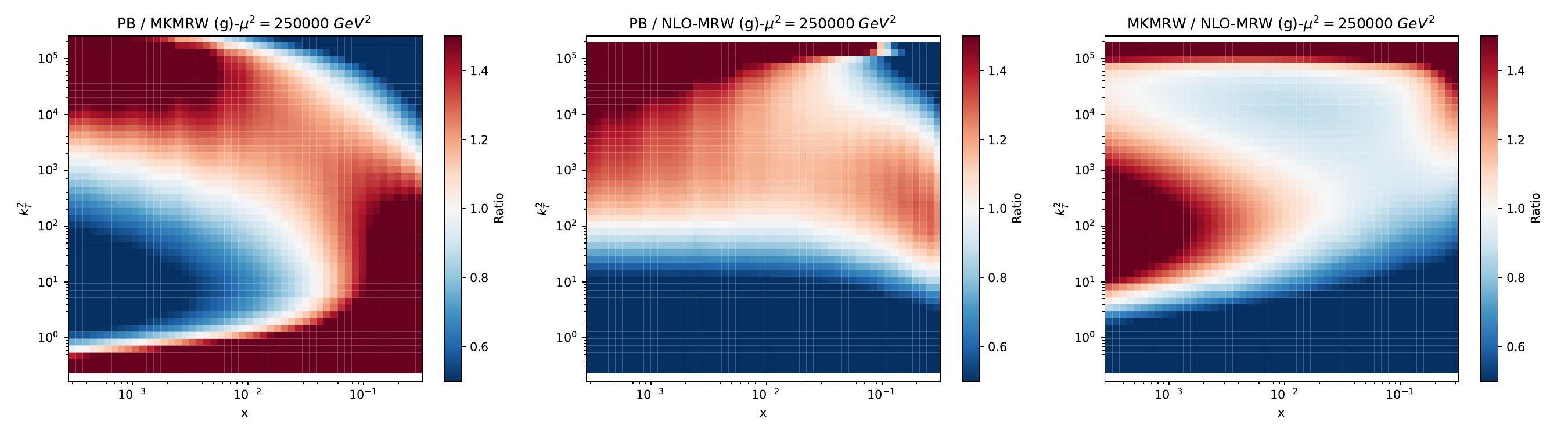}
	\caption{Comparison of different gluon UPDF models at $\mu = 70~\mathrm{GeV}$ (top) and 
		$\mu = 500~\mathrm{GeV}$ (bottom). The colour scale shows the ratio of UPDFs as a 
		function of the longitudinal momentum fraction $x$ of the gluon (horizontal axis) and 
		its transverse momentum $k_T$ in GeV (vertical axis), both on logarithmic scales.
		From left to right the panels display the ratios 
		$PB/MKMRW$, $PB/NLO\text{-}MRW$, and $MKMRW/NLO\text{-}MRW$}
	\label{fig:g_updfHeatmap}
\end{figure}

\begin{figure}[htbp]
	\centering
	\includegraphics[width=16cm, height=7.1cm]{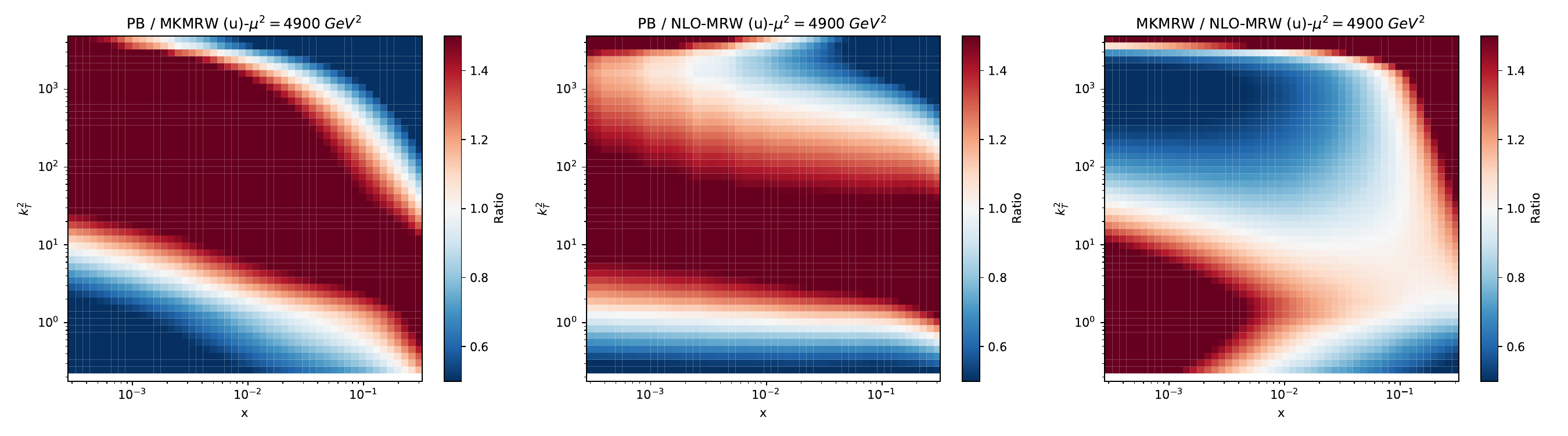}
	\includegraphics[width=16cm, height=7.1cm]{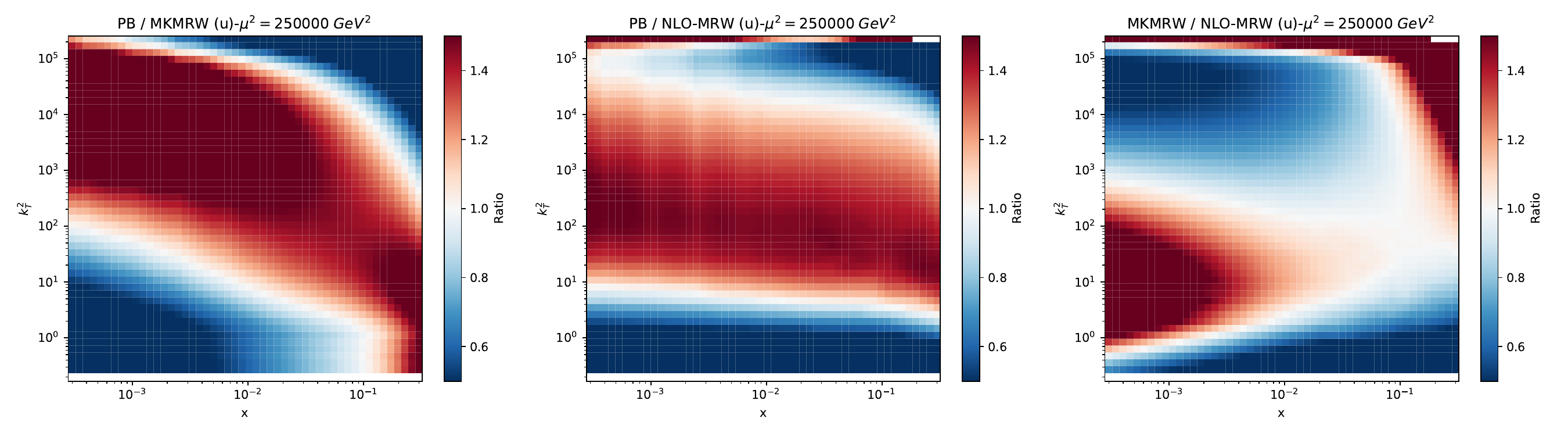}
	\caption{
		Comparison of different up-quark UPDF models at $\mu = 70~\mathrm{GeV}$ (top) and 
		$\mu = 500~\mathrm{GeV}$ (bottom). The colour scale shows the ratio of UPDFs as a 
		function of the longitudinal momentum fraction $x$ of the up quark (horizontal axis) 
		and its transverse momentum $k_T$ in GeV (vertical axis), both on logarithmic scales.
		From left to right the panels display the ratios 
		$PB/MKMRW$, $PB/NLO\text{-}MRW$, and $MKMRW/NLO\text{-}MRW$.}
	\label{fig:u_updfHeatmap}
\end{figure}
\begin{itemize}
	\item \textbf{Transverse momenta of photons:}  
	The transverse momentum $p_{T,\gamma_1}$ ($p_{T,\gamma_2}$) of the leading (sub-leading) photon.
	
	\item \textbf{Invariant mass  and transverse momentum of the diphoton system:}  
	The invariant mass of the diphoton is defined as
	\[
	m_{\gamma\gamma} = \sqrt{(E_{\gamma_1} + E_{\gamma_2})^2 - (\vec{p}_{\gamma_1} + \vec{p}_{\gamma_2})^2},
	\]
	and the transverse momentum of the diphoton system is given by the vector sum of the photon transverse momenta:
	\[
	\vec{p}_{T,\gamma\gamma} = \vec{p}_{T,\gamma_1} + \vec{p}_{T,\gamma_2}.
	\]
	
	\item \textbf{Scattering angle in the Collins–Soper frame:}  
	The absolute value of the cosine of the scattering angle with respect to the $z$-axis in the Collins–Soper frame is expressed as:
	\[
	\left|\cos \theta^{*}_{\mathrm{(CS)}}\right| = 
	\left|\frac{\sinh(\Delta\eta_{\gamma\gamma})}{\sqrt{1+(p_{T,\gamma\gamma}/m_{\gamma\gamma})^2}}
	\cdot \frac{2 p_{T,\gamma_1}p_{T,\gamma_2}}{m_{\gamma\gamma}^2}\right|
	= \left|\frac{p_{\gamma_1}^{+}p_{\gamma_2}^{-} - p_{\gamma_1}^{-}p_{\gamma_2}^{+}}{m_{\gamma\gamma}\sqrt{m_{\gamma\gamma}^2 + p_{T,\gamma\gamma}^2}}\right|,
	\]
	where $\Delta\eta_{\gamma\gamma} = \eta_{\gamma_1} - \eta_{\gamma_2}$ and $p_{\gamma}^{\pm} = E_{\gamma} \pm p_{z,\gamma}$. 
	This variable allows for a direct interpretation of the scattering angle in the presence of initial-state radiation.
	
	\item \textbf{Angular variable $\boldsymbol{\phi^{*}_{\eta}}$:}  
	An angular observable sensitive to $p_{T,\gamma\gamma}$, defined as:
	\[
	\phi^{*}_{\eta} = \tan\left(\frac{\pi - \Delta\phi_{\gamma\gamma}}{2}\right) 
	\sin\theta^{*}_{\eta} = 
	\tan\left(\frac{\pi - \Delta\phi_{\gamma\gamma}}{2}\right) 
	\sqrt{1 - \tanh^2\left(\frac{\Delta\eta_{\gamma\gamma}}{2}\right)}.
	\]
	This variable is particularly useful at low $p_{T,\gamma\gamma}$, where it provides a more precise probe of QCD dynamics owing to its superior angular resolution compared with direct $p_{T,\gamma\gamma}$ measurements.
	
	\item \textbf{Acoplanarity:}  
	The acoplanarity of the two photons is defined as $\pi - \Delta\phi_{\gamma\gamma}$, where $\Delta\phi_{\gamma\gamma}$ is their azimuthal angular separation.
	
	\item \textbf{Transverse thrust-related variable $\boldsymbol{a_{T,\gamma\gamma}}$:}  
	The transverse component of $p_{T,\gamma\gamma}$ with respect to the thrust axis is given by:
	\[
	a_{T,\gamma\gamma} = 2 \cdot 
	\frac{|p_{x,\gamma_1}p_{y,\gamma_2} - p_{y,\gamma_1}p_{x,\gamma_2}|}
	{|\vec{p}_{T,\gamma_1} - \vec{p}_{T,\gamma_2}|},
	\]
	where $p_{x(y),\gamma_{1(2)}}$ are the $x(y)$ components of the photon transverse momenta. 
	This observable, like $\phi^{*}_{\eta}$, is highly sensitive to low-$p_{T,\gamma\gamma}$ dynamics.
\end{itemize}

Before discussing and comparing the results obtained with different UPDF models and their
impact on the predicted differential cross sections, we briefly comment on the regions of
the unintegrated phase space that are most relevant to this study. For this purpose, the
Les Houches Event (LHE) files generated by the \textsc{KaTie} parton-level event generator
are analysed using the \texttt{pylhe} library~\cite{pylhe}. We study the differential cross
section as a function of the longitudinal momentum fractions $x_1$ and $x_2$ of the
incoming partons (moving along the $+\hat{z}$ and $-\hat{z}$ directions, respectively), the
transverse momenta $k_{t1}$ and $k_{t2}$ of the initial-state partons, and the factorization
scale $\mu_{\mathrm{fac}}$ of the subprocess.

From this analysis we find that, under the ATLAS fiducial cuts, the dominant contribution
to the diphoton cross section arises from the ranges
$3\times10^{-4}\lesssim x_{1,2}\lesssim 3\times10^{-1}$,
$0.2~\mathrm{GeV}\lesssim k_{t1,2}\lesssim 5~\mathrm{GeV}$ and
$\mu_{\mathrm{fac}}\sim 70$–$150~\mathrm{GeV}$. We have verified that the representative
subprocesses $q+\overline{q}\to\gamma+\gamma$ and $q+g\to\gamma+\gamma+q$ populate very
similar regions of this unintegrated phase space; other channels behave analogously.
These ranges are motivated the choices of scales used in the UPDF comparisons shown
in Figs.~\ref{fig:g_updfHeatmap} and~\ref{fig:u_updfHeatmap}.
\begin{figure}[htbp]
	\centering
	\includegraphics[width=7.5cm, height=7.1cm]{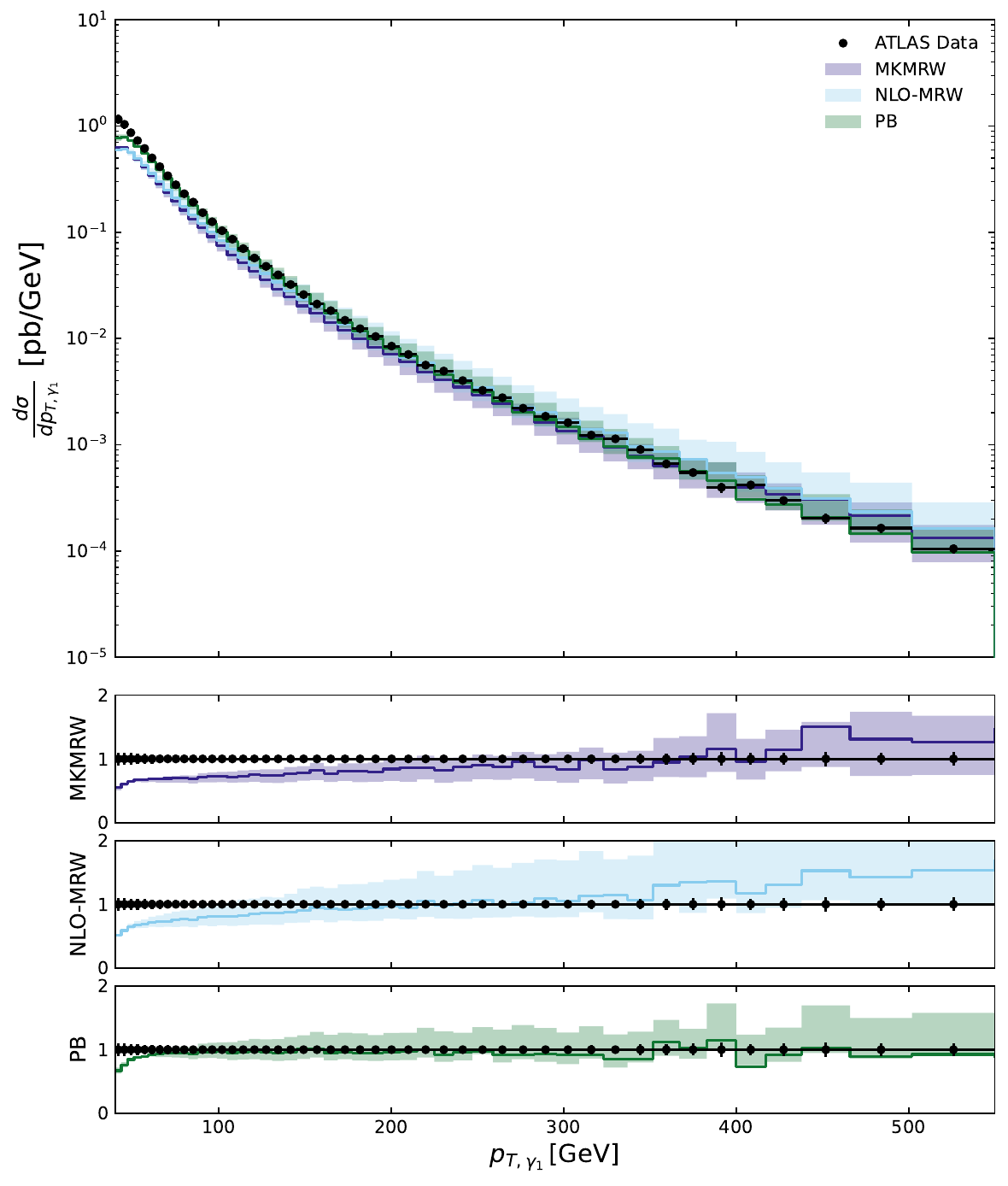}
	\includegraphics[width=7.5cm, height=7.1cm]{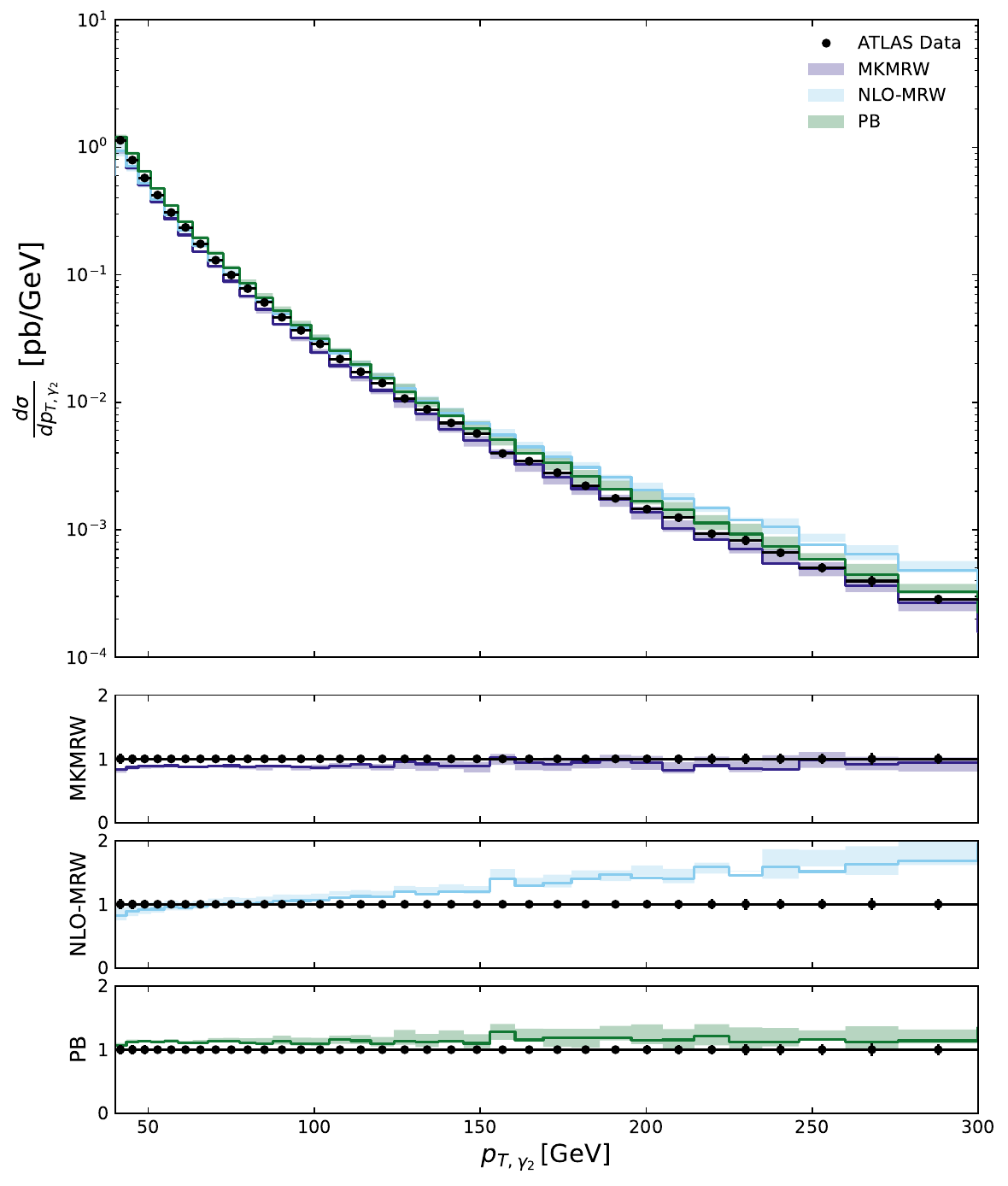}
	\caption{
		Comparison of the theoretical predictions from the NLO-MRW, MKMRW, and PB UPDF models with the experimental data for the transverse momenta of the two final-state photons. 
		The left panel shows the distribution for the leading photon, while the right panel corresponds to the subleading photon.
	}
	\label{fig:pt1_2photon}
\end{figure}

In the figure \ref{fig:g_updfHeatmap}, a comparison between different UPDF models is presented for $\mu = 70~\mathrm{GeV}$ (first row) and $\mu = 500~\mathrm{GeV}$ (second row), corresponding to the ratios $PB/MKMRW$, $PB/NLO-MRW$, and $MKMRW/NLO-MRW$, respectively. 
It is evident from this figure that, in general, the NLO-MRW model predicts considerably larger transverse momentum distributions than the PB and MKMRW models. 
This behaviour primarily arises from the scale choice $k^2 = k_t^2/(1-z)$ in the input PDFs of the NLO-MRW formulation. 
Consequently, at small transverse momenta, the NLO-MRW model effectively probes PDFs at higher scales, which naturally leads to larger UPDF values in this region. 
Interestingly, all UPDF models yield relatively similar results in the medium-$k_t$ region. 
Another noteworthy feature, as will be discussed later, is that for large $x$ and some moderate-to-high $k_t$ regions, the NLO-MRW UPDF model predicts larger distributions, whereas for $k_t$ values approaching the factorization scale, it is suppressed by the strong virtuality-order cutoff $\Theta(\mu^2 - k^2)$ intrinsic to the model. Additionally, the MKMRW and NLO-MRW UPDF models exhibit relatively similar behaviour; thus, comparable results are expected from these two formulations. 
The PB model, however, generally predicts larger UPDF values in the medium-$k_t$ region compared with the others.
\begin{figure}[htbp]
	\centering
	\includegraphics[width=7.5cm, height=7.1cm]{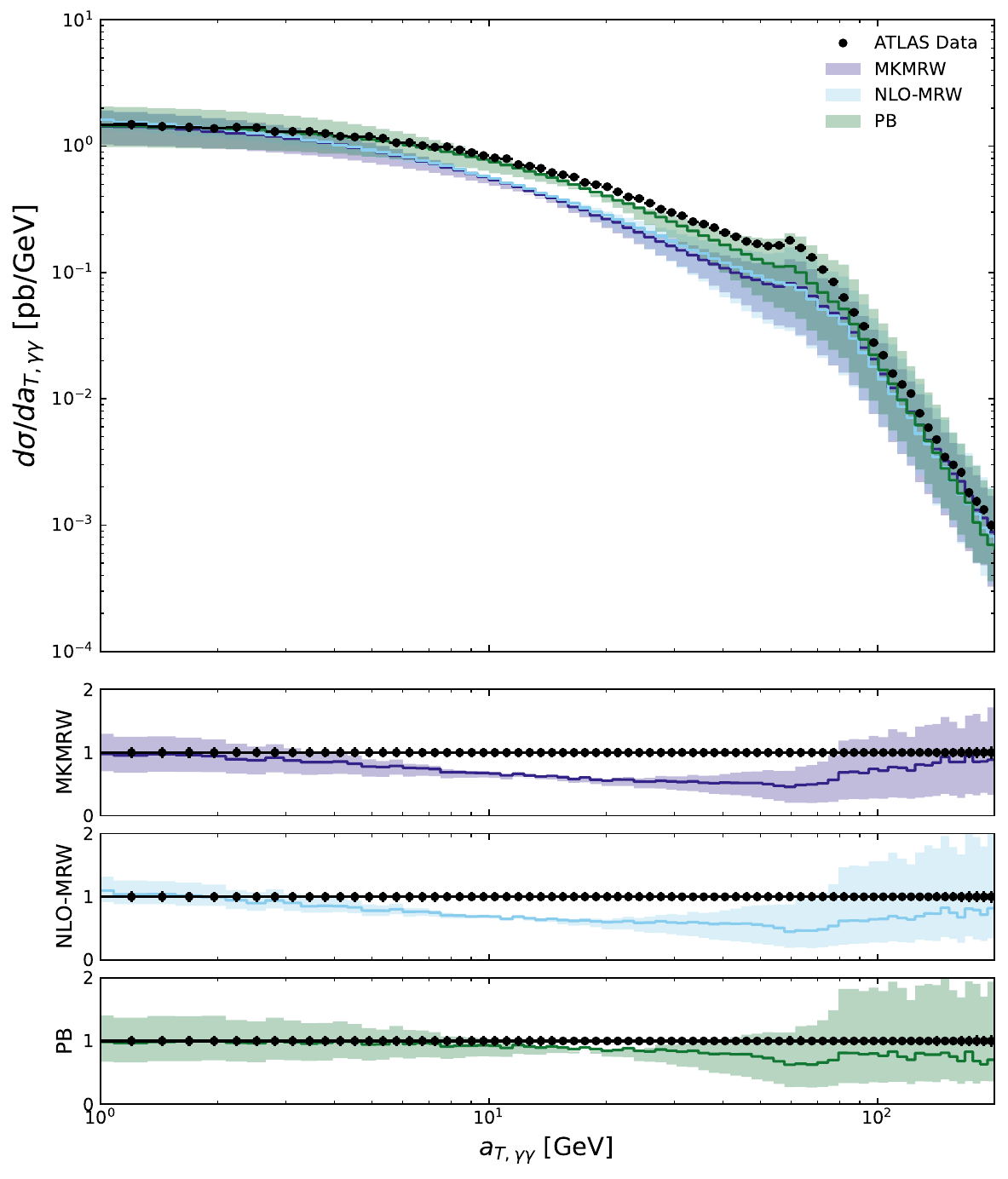}
	\includegraphics[width=7.5cm, height=7.1cm]{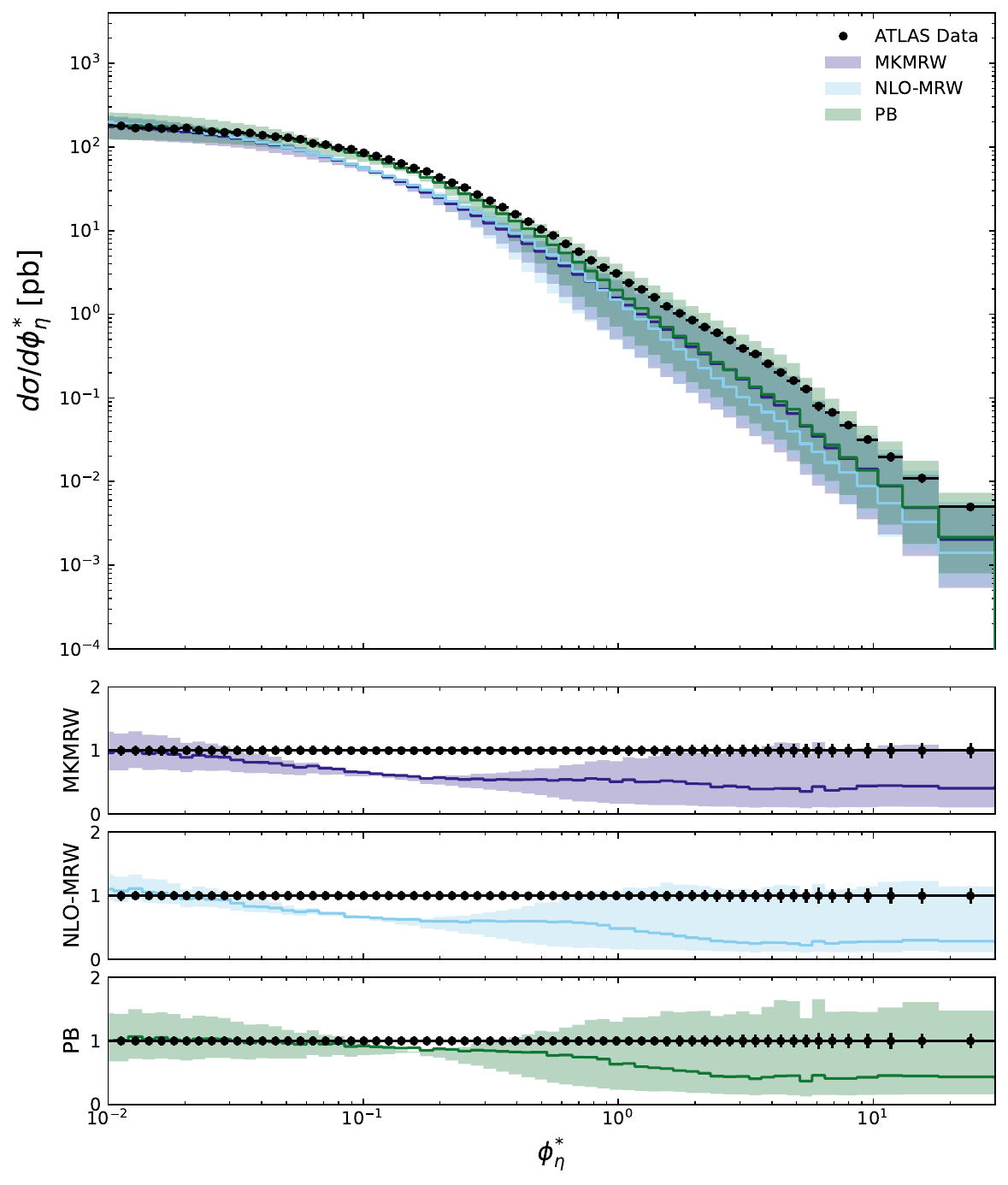}
	\includegraphics[width=7.5cm, height=7.1cm]{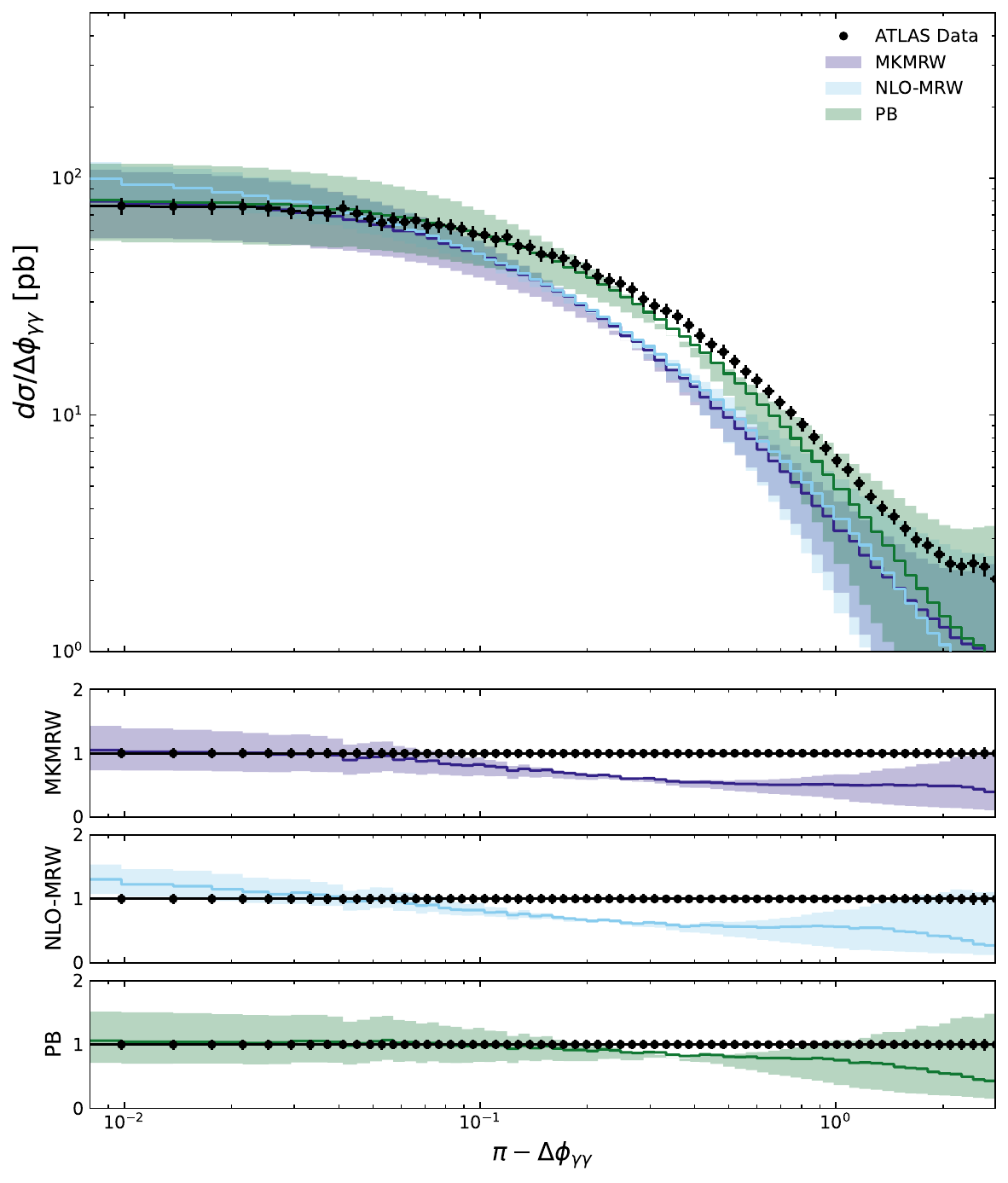}
	\includegraphics[width=7.5cm, height=7.1cm]{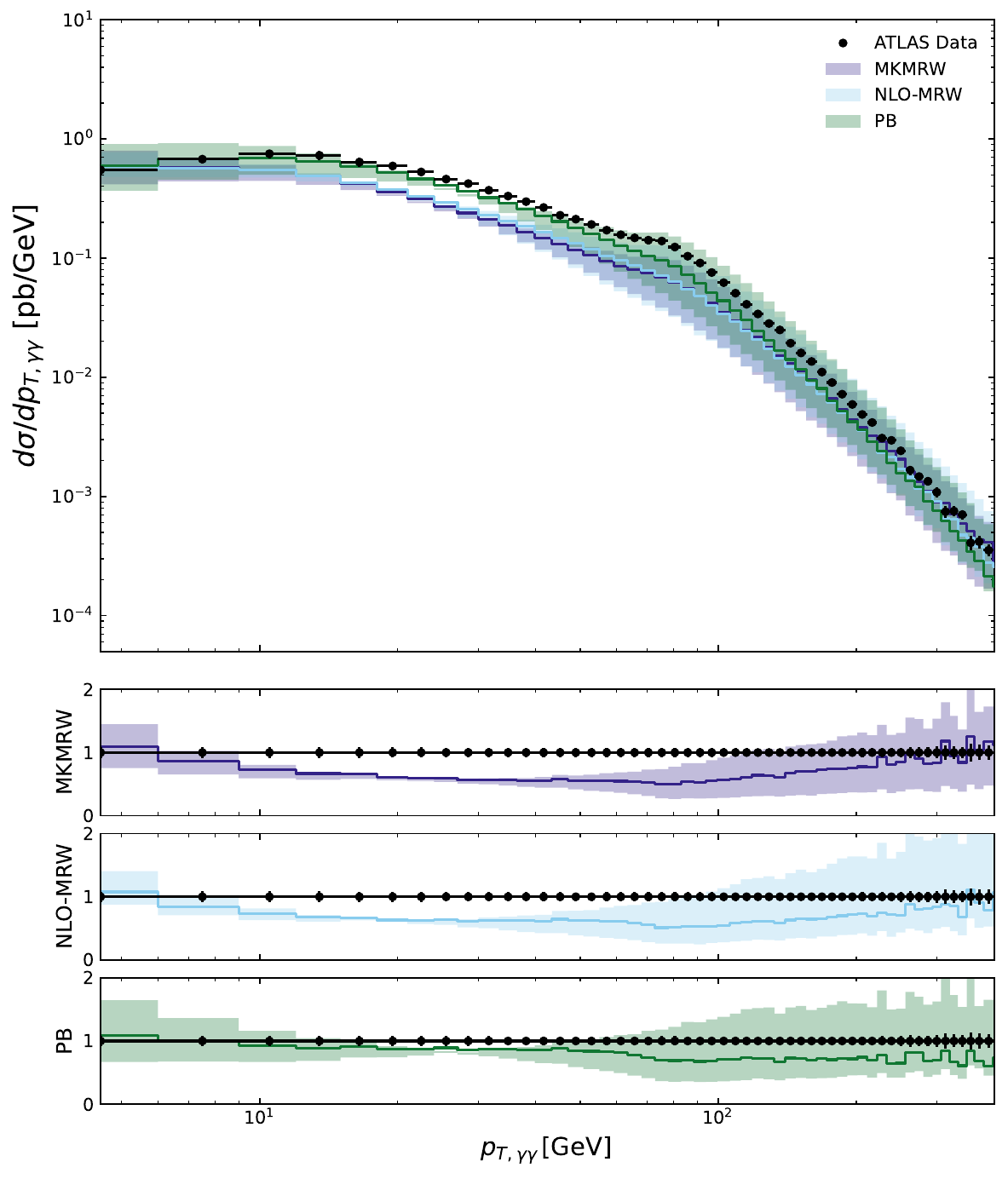}
	\caption{
		Comparison of the theoretical predictions from the NLO-MRW, MKMRW, and PB UPDF models with experimental data for four observables: 
		(top-left) $a_{T,\gamma\gamma}$, (top-right) $\phi^{*}_{\eta}$, (bottom-left) $\pi - \Delta\phi_{\gamma\gamma}$, and (bottom-right) $p_{T,\gamma\gamma}$. 
		The PB model provides the best overall agreement with the experimental data across all observables, 
		while the MKMRW and NLO-MRW models tend to underestimate the data in the intermediate regions but reproduce the measurements within uncertainties at low and high values.}
	\label{fig:atAA_phistar_piMinDPhi_ptAA}
\end{figure}
\begin{figure}[htbp]
	\centering
	\includegraphics[width=7.5cm, height=7.1cm]{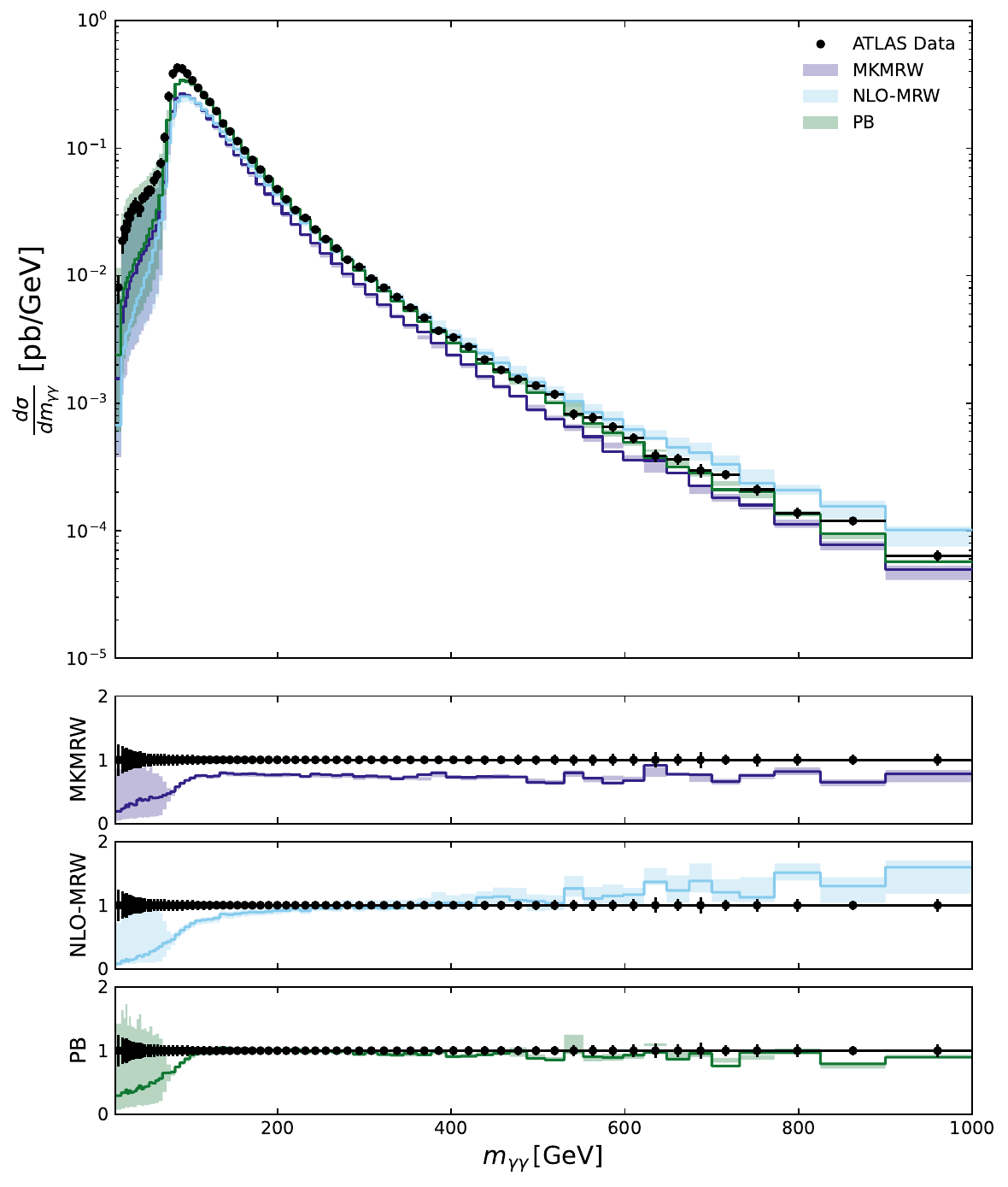}
	\includegraphics[width=7.5cm, height=7.1cm]{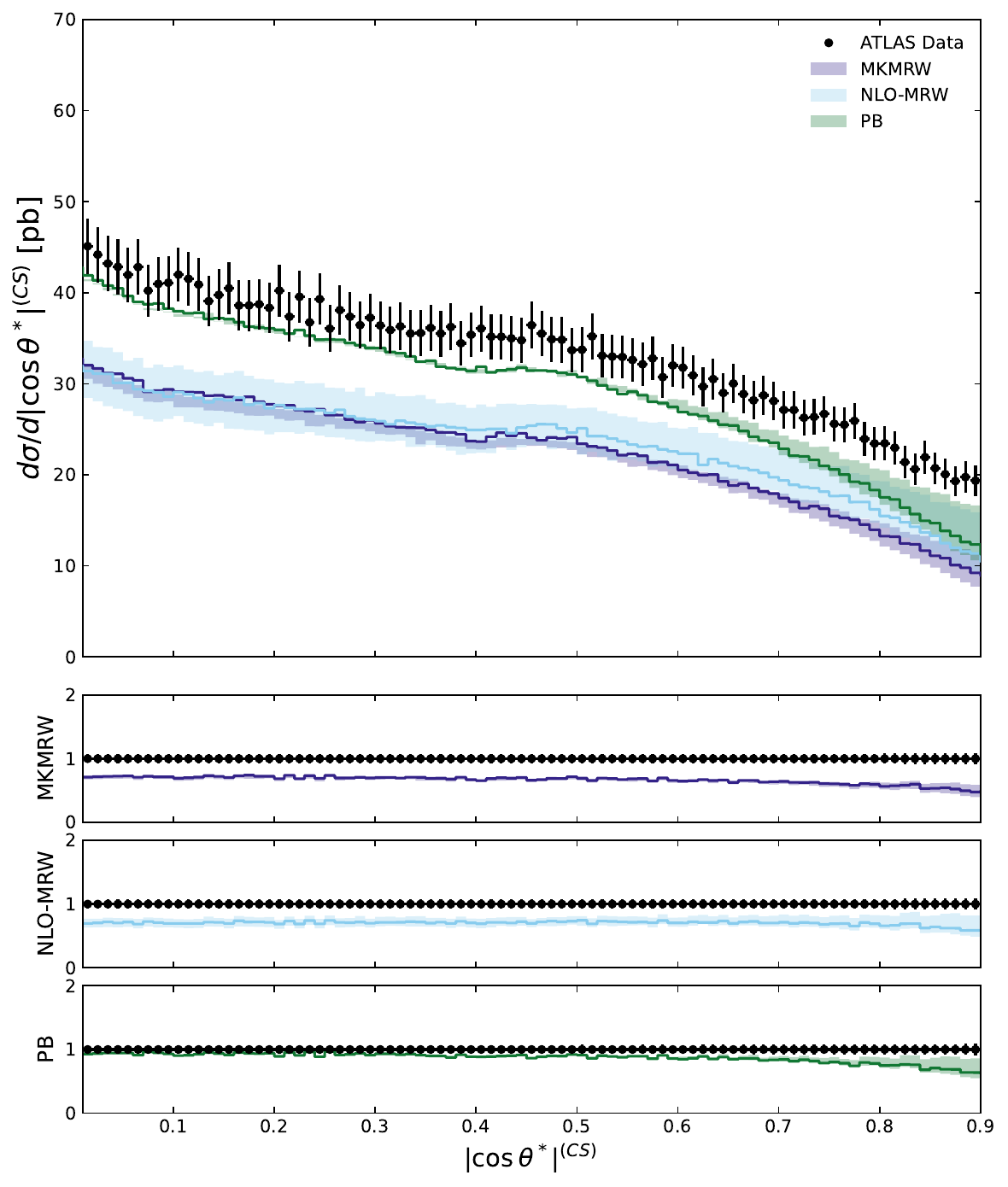}
	\caption{
		Comparison of the theoretical predictions from the NLO-MRW, MKMRW, and PB UPDF models with experimental data for 
		(left) the diphoton invariant mass $m_{\gamma\gamma}$ and (right) the Collins–Soper angle $\left|\cos \theta^{*}_{\mathrm{(CS)}}\right|$. 
		The PB model provides the best overall agreement with the data across both observables, 
		whereas the NLO-MRW and MKMRW models tend to overestimate and underestimate the data, respectively, 
		particularly in regions sensitive to higher factorization scales.}
	\label{fig:mAA_CosThetaStar}
\end{figure} 
Finally, the MKMRW distributions are noticeably smaller than those of the other models, particularly in regions with very low parton transverse momenta. 
This suppression arises from the choice of an infinite upper limit in the Sudakov form factor.

In the figure \ref{fig:u_updfHeatmap}, a similar comparison is shown for the up-quark UPDFs. 
The overall patterns are consistent with those observed for gluons; however, the PB UPDF tends to predict larger up-quark densities, particularly in the medium-$k_t$ region.

Before presenting our predictions for the measured observables, we note that the ATLAS data points shown in this section are taken from the HEPData tables corresponding to 
Ref.~\cite{ATLAS:2021mbt}. For each bin, the experimental uncertainty shown in our figures is the total experimental uncertainty, obtained by combining in quadrature all individual 
components listed in HEPData (statistical and systematic, with upward and downward variations treated separately).

In the left and right panels of the figure \ref{fig:pt1_2photon}, we compare the predictions of the NLO-MRW, MKMRW, and PB UPDF models with the experimental data. 
Our analysis indicates that these two differential cross sections are strongly correlated with the factorization-scale dependence of the UPDFs. 
As evident from the figure \ref{fig:g_updfHeatmap} and figure \ref{fig:u_updfHeatmap}, increasing the factorization scale causes the NLO-MRW UPDFs to become significantly larger than those of the other models. 
Consequently, it is expected that at higher transverse momenta of the final-state photons, the NLO-MRW model overestimates the corresponding predictions of the other UPDF frameworks. 

Furthermore, these figures show that while the NLO-MRW model provides a reasonable description of the transverse momentum distribution of the leading photon (the photon with the highest $p_{T}$), it fails to reproduce the experimental data at large $p_{T}$ values for the subleading photon (the photon with the second-highest $p_{T}$). 
In contrast, both the MKMRW and PB models yield predictions that are in better agreement with the experimental measurements across most of the kinematic range.

In the top-left and top-right panels of the figure \ref{fig:atAA_phistar_piMinDPhi_ptAA}, we compare the differential cross sections with respect to $a_{T,\gamma\gamma}$ and $\phi^{*}_{\eta}$, respectively, using predictions from different UPDF models. 
Additionally, the bottom-left and bottom-right panels of this figure present similar comparisons for $\pi - \Delta\phi_{\gamma\gamma}$ and $p_{T,\gamma\gamma}$, respectively.

These four observables are predominantly correlated with the transverse momenta of the incoming partons. 
As discussed previously and illustrated in the figure \ref{fig:g_updfHeatmap} and figure \ref{fig:u_updfHeatmap}, the PB UPDFs generally tend to overestimate the other models in the medium-to-large transverse momentum region, whereas the NLO-MRW model predicts considerably larger values at relatively small transverse momenta. 

It can be observed that among the different UPDF models, only the PB model provides a consistent and accurate description of the experimental data across all four observables, $a_{T,\gamma\gamma}$, $\phi^{*}_{\eta}$, $\pi - \Delta\phi_{\gamma\gamma}$, and $p_{T,\gamma\gamma}$. 
In contrast, both the MKMRW and NLO-MRW models tend to undershoot the data in the intermediate regions. 
Nevertheless, the MKMRW and NLO-MRW predictions remain compatible with the experimental measurements at low and high values of these observables once uncertainties are taken into account. Additionally, it should be mentioned that within the collinear factorization framework, fixed-order calculations are known to be unreliable in the small-$a_{T,\gamma\gamma}$, small-$\phi^{*}_{\eta}$, small-$(\pi-\Delta\phi_{\gamma\gamma})$, and small-$p_{T,\gamma\gamma}$ regions, i.e. where parton-shower or resummation effects become essential~\cite{ATLAS:2021mbt}. In the $k_t$-factorization approach, the Sudakov form factor embedded in the UPDFs effectively resums soft and collinear emissions, yielding a more stable and broadly consistent description of these regions.

In the left and right panels of the figure \ref{fig:mAA_CosThetaStar}, we compare the differential cross sections with respect to the diphoton invariant mass $m_{\gamma\gamma}$ and the Collins–Soper angle $\left|\cos \theta^{*}_{\mathrm{(CS)}}\right|$, respectively, for different UPDF models. 
It should first be noted that $m_{\gamma\gamma}$ is strongly correlated with the factorization scale. 
As a result, the NLO-MRW UPDF model, similar to its behaviour in the differential cross sections with respect to the transverse momenta of the leading and subleading photons, tends to overestimate the experimental data at large diphoton invariant masses. 
In contrast, the MKMRW model generally underestimates the data, while the PB approach successfully reproduces the experimental measurements across the entire kinematic range. 

An interesting feature appears around the ``shoulder'' region, where $m_{\gamma\gamma} \approx \min(p_{T,\gamma_1}) + \min(p_{T,\gamma_2})$. 
In this region, only the PB model provides an accurate description of the experimental data, demonstrating its superiority over the other UPDF models. 
This superiority becomes even more evident when comparing the model predictions for $\left|\cos \theta^{*}_{\mathrm{(CS)}}\right|$. 
Among the different UPDF models, only the PB approach yields a consistent and reliable description of the experimental data across all regions.

Overall, the comparison of the various UPDF models with both the experimental data and each other reveals that the PB approach provides the most accurate and stable predictions over the full kinematic range. 
The NLO-MRW model tends to overestimate the data, particularly in regions and observables sensitive to larger factorization scales, while the MKMRW model generally exhibits similar trends but with systematically lower predictions due to the stronger suppression effects from the Sudakov form factor, whose upper integration limit is set to infinity.

\section{Conclusion}

In this work, we have investigated isolated prompt diphoton production in proton–proton collisions at $\sqrt{s}=13~\mathrm{TeV}$ within the $k_t$-factorization framework. The analysis incorporated several UPDF models (namely, the PB, NLO-MRW, and MKMRW approaches) and compared their predictions with ATLAS experimental data. The relevant partonic subprocesses were evaluated using the \textsc{KaTie} event generator for tree-level contributions and an independent implementation for the loop-induced $g + g\to\gamma + \gamma$ channel.  

A comprehensive set of kinematic observables was analyzed, including photon transverse momenta, diphoton invariant mass and transverse momentum, angular correlations, and observables sensitive to soft radiation such as $\phi^{*}_{\eta}$ and $a_{T,\gamma\gamma}$. The results consistently show that the PB UPDF model provides the best overall agreement with the experimental data across all measured distributions. In particular, the PB formalism successfully reproduces the shape and normalization of the data over both low and high-$p_T$ regions, reflecting its realistic treatment of parton transverse momenta and coherence effects during evolution.

 In contrast, the NLO-MRW model tends to overestimate the cross sections in regions dominated by large factorization scales, a consequence of the $k^2 = k_t^2/(1-z)$ scale choice in its formulation. The MKMRW approach, while theoretically improved through its modified Sudakov form factor and extended normalization, generally underestimates the data due to stronger suppression at large $k_t$.

Overall, our findings confirm that the $k_t$-factorization approach provides a reliable framework for describing photon pair production at the LHC. The PB formalism, in particular, emerges as the most accurate and stable UPDF model currently available, combining perturbative consistency with excellent phenomenological performance.

\section*{Acknowledgements}
We would like to express our sincere gratitude to the Institute for Research in Fundamental Sciences (IPM) for their financial support and research facilities. 
We also extend our appreciation to Dr.~Benjamin Guiot for his valuable comments and guidance on the implementation of the MKMRW UPDF model.

\bibliographystyle{apsrev4-2}
\bibliography{references} 
\end{document}